# Using game simulator Software Inc in the Software Engineering education

Tetiana A. Vakaliuk[1][0000-0001-6825-4697], Valerii V. Kontsedailo[1][0000-0002-6463-370X],
Dmytro S. Antoniuk[1][0000-0001-7496-3553], Olha V. Korotun[1][0000-0003-2240-7891],
Iryna S. Mintii[2][0000-0003-3586-4311] and Andrey V. Pikilnyak[3][0000-0003-0898-4756]

[1] Zhytomyr Polytechnyc State University, 103, Chudnivska Str., Zhytomyr, 10005, Ukraine
[2] Kryvyi Rih State Pedagogical University, 54, Gagarin Ave., Kryvyi Rih, 50086, Ukraine
[3] Kryvyi Rih National University, 11, Vitali Matusevich St., Kryvyi Rih, 50027, Ukraine
tetianavakaliuk@gmail.com, valerakontsedailo@gmail.com,
dmitry_antonyuk@yahoo.com, olgavl.korotun@gmail.com,
irina.mintiy@kdpu.edu.ua, pikilnyak@gmail.com

**Abstract.** The article presents the possibilities of using game simulator Sotware Inc in the training of future software engineer in higher education. Attention is drawn to some specific settings that need to be taken into account when training in the course of training future software engineers. More and more educational institutions are introducing new teaching methods, which result in the use of engineering students, in particular, future software engineers, to deal with real professional situations in the learning process. The use of modern ICT, including game simulators, in the educational process, allows to improve the quality of educational material and to enhance the educational effects from the use of innovative pedagogical programs and methods, as it gives teachers additional opportunities for constructing individual educational trajectories of students. The use of ICT allows for a differentiated approach to students with different levels of readiness to study. A feature of any software engineer is the need to understand the related subject area for which the software is being developed. An important condition for the preparation of a highly qualified specialist is the independent fulfillment by the student of scientific research, the generation, and implementation of his idea into a finished commercial product. In the process of research, students gain knowledge, skills of the future IT specialist and competences of the legal protection of the results of intellectual activity, technological audit, marketing, product realization in the market of innovations. Note that when the real-world practice is impossible for students, game simulators that simulate real software development processes are an alternative.

**Keywords:** simulator, game simulator, training, software engineers.

More and more educational institutions are introducing new teaching methods, which result in the use of engineering students, in particular, majoring in software engineering, to deal with real professional situations in the learning process [30, p. 547; 41, p. 150].

The use of modern ICT, including game simulators, in the educational process [31; 37; 13; 19; 24; 29; 60], allows to improve the quality of educational material and to enhance the educational effects from the use of innovative pedagogical programs and methods, as it gives teachers additional opportunities for constructing individual educational trajectories of students [55]. The use of ICT allows for a differentiated approach to students with different levels of readiness to study.

A feature of any software engineer is the need to understand the related subject area for which the software is being developed [57]. An important condition for the preparation of a highly qualified specialist is the independent fulfillment by the student of scientific research, the generation, and implementation of his idea into a finished commercial product. In the process of research, students gain knowledge and skills of the future IT specialist among with the competences of the legal protection of the results of intellectual activity, technological audit, marketing, product realization in the market of innovations. Note that when the real-world practice is impossible for students, game simulators that simulate real software development processes are an alternative.

This topic is dedicated to the research of Liudmyla I. Bilousova [6], Yevhenii O. Modlo [39], Ivan O. Muzyka [28], Pavlo P. Nechypurenko [38], Olga P. Pinchuk [50], Andrii M. Striuk [61], and others. In particular, computerization and informatization of the education system were considered by Oleksandr Yu. Burov [50], Arnold E. Kiv [25], Oleksandr H. Kolgatin [27], Maiia V. Marienko [51], Liubov F. Panchenko [48], Svitlana V. Shokaliuk [35], Vladimir N. Soloviev [17], Tetiana V. Starova [44], Viktoriia V. Tkachuk [59], Snizhana O. Zelinska [58], and others; Olga V. Bondarenko [7], Svitlana H. Lytvynova [32], Pavlo P. Nechypurenko [43], Yuliya H. Nosenko [46], Maiia V. Marienko [33], Oksana M. Markova [45], Maryna V. Rassovytska [53], Serhiy O. Semerikov [40], Mariya P. Shyshkina [56], Andrii M. Striuk [34], and others, paid attention to the use of electronic learning tools.

Importance and necessity of introduction of information and communication technologies (ICT), including game simulators, in training are substantiated by Faheem Ahmed [1], Ritika Atal [4], Alex Baker [5], Bassam Baroudi [49], Márcio Barros [12], Kunal Bedse [23], Salah Bouktif [1], Noel Burchell [20], Alejandro Calderón [9], Piers Campbell [1], Luiz Fernando Capretz [1], Craig Caulfield [10], Paul Clarke [11], Alexandre Dantas [12], Khaled El Emam [15], Ayesha Farooq [52], Dave Hodges [20], André van der Hoek [5], Mehdi Jazayeri [22], Hashem Salarzadeh Jenatabadi [47], Shanika Karunasekera [23], A. Güneş Koru [15], Stanislaw Paul Maj [10], Emily Oh Navarro [42], Ali Noudoostbeni [47], Rory V. O'Connor [11], Ira Pant [49], Sanghamitra Patnaik [52], Goparaju Purna Sudhakar [52], Lise Renaud [54], Mercedes Ruiz [9], Louise Sauvé [54], Ashish Sureka [4], David Veal [10], Cláudia Werner [12], Jianhong Xia [10], Norizan Mohd Yasin [47]. ICTs are part of every area of human activity and have a positive impact on education, as they open up opportunities for the introduction of completely new teaching and learning methods.

A significant contribution to the theory of educational games was made by Daniil B. Elkonin [16], Eric Klopfer [26], David R. Michael [36], Lev S. Vygotskii [63], and others. At the same time, game technologies of teaching and use of interactive games in high school were investigated by Ayşe Alkan [2], Mohammad Hasan Al-Tarawneh [3], Viktoriia L. Buzko [8], Muhammet Demirbilek [14], Glenda A. Gunter [18],

G. Tanner Jackson [21], Estela Aparecida Oliveira Vieira [62], Nataliia P. Volkova [60], and other modern scholars and pedagogues. However, the question of the use of game simulators in the training of future software engineers remains poorly understood.

That is why the purpose of this article is to describe the features of working with the game simulator Software Inc in the software engineering training.

Game simulators are interactive programs that fully or partially simulate certain real processes or systems that capture and motivate students through fun and interesting game experiences, where students can perform different roles in a variety of realistic circumstances and are used in the educational process when real practice is impossible or inaccessible.

Software Inc is a game simulator that allows students to try their hand at running a software development company.

To start the game you need to press the "New Game" button, then the corresponding screen, shown in Fig. 1.

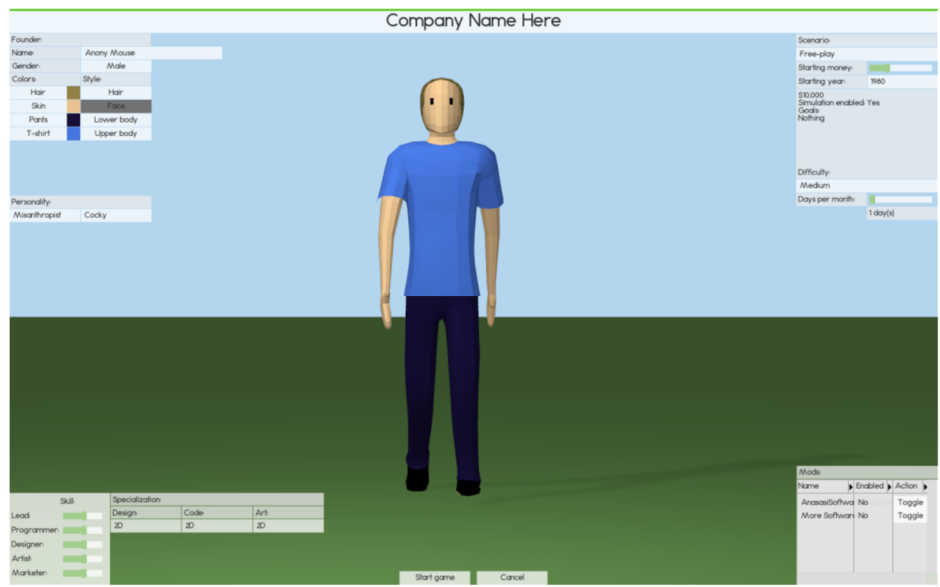

**Fig. 1.** Starting a game simulation in Software Inc

In the first stages, students should use a manual to provide detailed information about the gameplay and the various individual stages of the game. Beginners are encouraged to choose Optimism and Generosity as the main features when creating a company founder, without necessarily changing the settings of the sliders located on the left in the user interface.

In the panel on the right, you need to increase the startup capital of the company to $ 20,000 or move the slider one mark if the game is in a different currency (the currency can be changed in the options menu). Set to default 1980 as the year the company was

founded, there is no need to change, but at the stage of getting acquainted with the game the difficulty level must be set to "Easy".

The Days per month parameter sets the number of days in a month. By default, it is set to 1 day, meaning one game day will count as a month. With this setting, you can change the speed of the game (most of them set 4 days, but first you need to set the value to 1 or 2 days).

After that, the student needs to change the founding person's settings and choose a name for their company, such as "SpaceTech" or "Aperture Cake Production", after which the students will see the next window (see Figure 2).

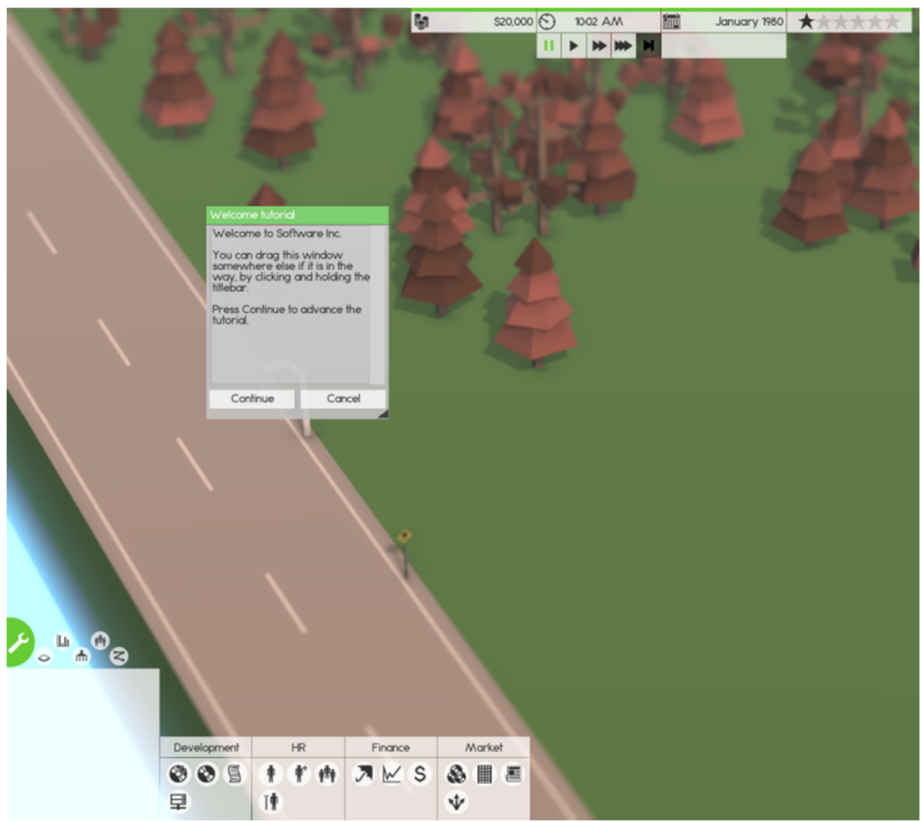

**Fig. 2.** Welcome dialog in Software Inc

You can move the instruction box around the screen and resize it. You have to press the "Continue" button to flip through the manual. To start creating the premises of a future virtual company for a student, it is necessary to go into construction mode by clicking on the green button with the image of a wrench on the user interface.

Game Simulator Software Inc offers three build modes:

− "Construct" – construction of rooms, installation of windows and doors;

─ "Furnish" – room furniture, choice of tables, chairs, coffee machine;
─ "Roads" – road construction, parking, but initially this mode is not used.

Once enabled, the student must first build a small room for the virtual company. To do this, you need to use the Room Builder tool in the Construct mode menu. The initial size of the first room should be approximately 5x5 to accommodate the table and chair for the founder of the company.

To make the room light enough, you must select the windows and doors in the Construct mode menu and install them in the room.

Then students have to go into Furnish mode and have a desk, chair, and computer installed in the room. Use Shift + left mouse button to set multiple items at once. Chairs will automatically be installed near the tables they are closest to, but you may need to turn your computer over.

Students need to click on and hold the computer, turning it in the required direction. After that, the room will look something like this (see Figure 3).

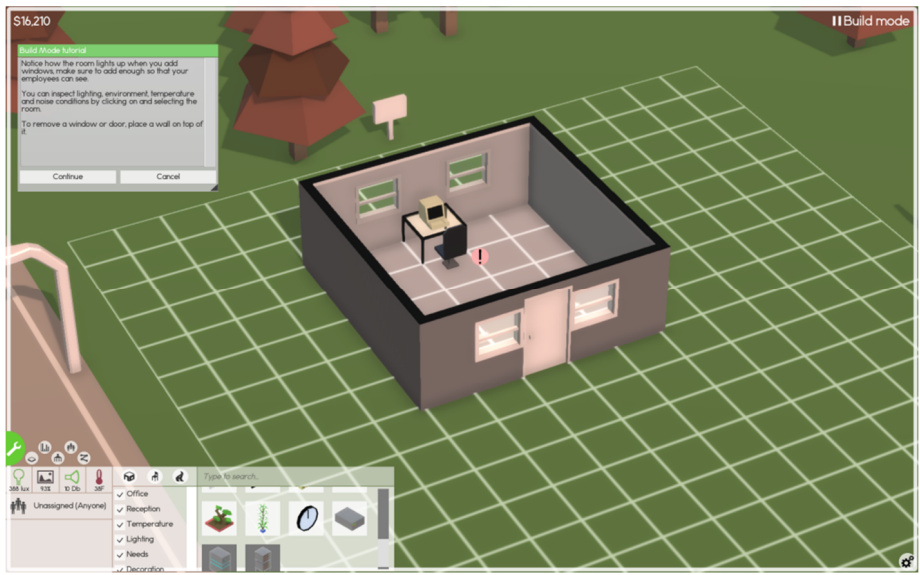

**Fig. 3.** A room built-in game simulator Software Inc

Once the room has been created, the founder of the company can start working there.
After reading the instructions, you must press the Tab key to exit the build mode.
*The process of making money and working on contracts.*
The first way that allows companies to make money is to work on contracts. Contracts are projects that are offered to companies for execution, and each of them has their requirements.

To find a contract, the participant must go to the "Development" menu by clicking the button with the image of a piece of paper (A). Then a pop-up window will appear on the screen (see Figure 4).

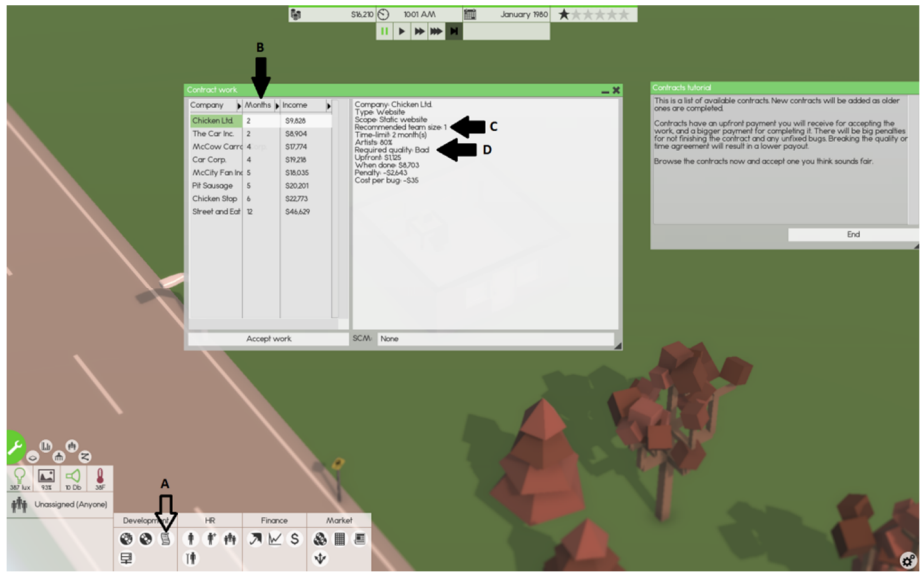

**Fig. 4.** Software search contract simulator popup window

Players need to sort the list by "Months" (B) to ensure that contracts that require the least number of months are at the top of the list. Typically, contracts with 1 or 2 (C) contracts will have a performance period of 1 or 2 months. The company does not receive big profits for working on such contracts, but in the beginning, it is a great source of financial income.

Regarding the quality of contract work, students must set the value to "Bad" or "Horrible" (D), which allows the initial stages not to take quality but quantity.

Students need to find a contract for their virtual company with a minimum work requirement for a team of 1 or 2 people for a maximum of two months and click the "Accept Work" button.

*Contract work.*

After the participants have selected a contract for their company, a project management pop-up window (A) will appear to the right in the user interface (see Figure 5). There are 4 stages of project implementation. The first stage is the development of the design. The product design specified in the contract is developed by the virtual company designer (s).

When pointing the mouse at the project management window, students can see the progress scale of a specific job by a company employee (B). The task is considered completed when the scale reaches its maximum value. However, students need to be careful and not delay the contracting process at the design development stage so as not to waste the time allotted for the contract.

Players must then click on the "Develop" (C) button in the project management window to proceed to the next stage – "Alpha". At this stage, the software engineers create the product according to the design developed by the designer in the first stage. Students can monitor the completion of the assignment using the progress bar. When the task is completed, you must click on the "Promote" button to proceed to the next step.

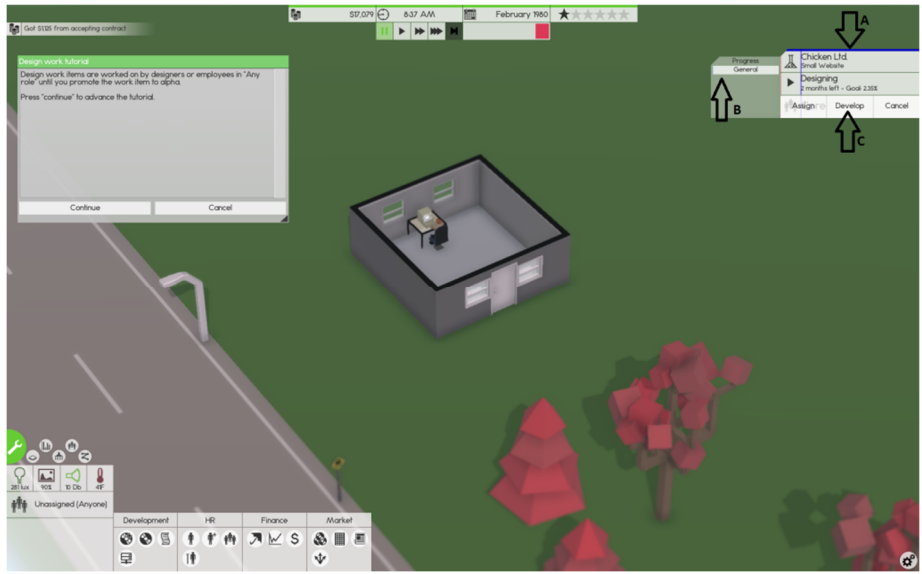

**Fig. 5.** The project management popup in the game simulator Software Inc

The next stage is called "Delay" and is only an intermediate stage, which lasts a certain amount of time depending on the skill level of the company's employees. The higher the skill level of the employees, the less time the Delay stage takes.

The project then proceeds to the "Beta" stage, which is to test and prepare for the product release. The company may release a product that is not ready for completion, but it is important to remember that the presence of errors and defects in the product will adversely affect its sales and contract payments.

*Building for employees and first employee.*

When building premises, students need to remember that employees are first and foremost people who have basic needs (want to eat, go to the restroom, drink coffee, prefer quiet, comfortable rooms with sufficient lighting, comfortable air temperature, and a comfortable environment). Therefore, you should build a facility that addresses all these needs, and do not forget to build small facilities for restrooms and rest areas.

Make sure you also have a fridge and coffee table in the lounge. Windows, lighting, and doors between rooms are required. It is important to keep in mind that indoor plants will have a positive effect on employee productivity and mood. Elements such as heat radiators and ventilation in the main room are also very important as they provide comfortable air temperatures.

How to hire employees?

After the company has reached a certain level of development, students need to hire new employees in the team.

When looking for a new employee, you first need to answer two questions:

1. In what position to open a vacancy?

- team lead;
- programmer;
- designer;
- creative manager;
- marketer.

2. How much time and resources can be spent on finding a new employee?

Then more time is spent in searching, then more applications from candidates will be considered. At the same time, it is necessary to consider the insurance policy of the company. Having a solid insurance fund will be an attractive factor for skilled workers. Therefore, every time after the completion of contract work and products, students need to invest in improving the insurance fund to attract highly qualified employees to the company.

After the list of candidates is created, you can go to the interviewing stage, during which you need to determine who should be hired and who should be better off. Keep in mind that you need to hire those who can be part of the team, as this will have a positive effect on the performance of the company and help you get things done faster.

The following is an example of a candidate whose compatibility with an existing team is low (see Figure 6).

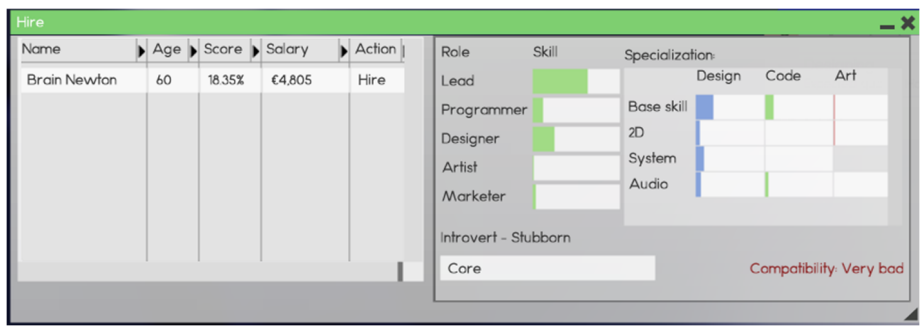

**Fig. 6.** Example of showing poor candidate compatibility with an existing team in a game simulator Software Inc

At the same time, there is a candidate whose compatibility with the team is sufficient (see Figure 7).

By finding a candidate with sufficient compatibility with the development team, as well as sufficient professional performance, players can hire a suitable candidate for their development team. They can then focus on managing software development or, if they see fit, continue to hire employees.

**Fig. 7.** An example of showing good candidate compatibility with an existing team in a simulator Software Inc